\documentclass[journal]{IEEEtran}
\usepackage{amsmath,amsfonts}
\usepackage{algorithm}
\usepackage{array}
\usepackage[caption=false]{subfig}
\usepackage{textcomp}
\usepackage{stfloats}
\usepackage{url}
\usepackage{verbatim}
\usepackage{graphicx}
\usepackage{cite}
\usepackage{verbatim} 
\usepackage{amsmath}
\usepackage{amssymb}
\usepackage{xspace} 
\usepackage{bm}
\usepackage{booktabs}
\usepackage{algpseudocode}
\usepackage{multirow}
\usepackage{changepage}
\usepackage{graphicx} 
\usepackage{tikz}
\usetikzlibrary{positioning,arrows.meta,calc,decorations.pathreplacing,fit,backgrounds}
\usetikzlibrary{positioning,fit,arrows.meta}
\usepackage{xcolor}
\usetikzlibrary{calc}
\usetikzlibrary{decorations.pathreplacing}
\usepackage[table,xcdraw,dvipsnames]{xcolor}

\definecolor{myNavy}{HTML}{001871}
\definecolor{myOrange}{HTML}{FFA300}
\definecolor{myRed}{HTML}{f9423a}
\definecolor{myBlue}{HTML}{00a9e0}
\definecolor{myGreen}{HTML}{78BE20}
\definecolor{myYellow}{HTML}{FFD100}
\definecolor{myPurple}{HTML}{70489e}

\usepackage{tikz}
\usetikzlibrary{positioning,fit,arrows.meta,calc}
\usepackage{xcolor}

\definecolor{stagegreen}{RGB}{120,190,32}
\definecolor{stageblue}{RGB}{0,181,226}
\definecolor{stagepeach}{RGB}{225,163,0} 
\definecolor{stagelilac}{RGB}{225,195,225} 
\definecolor{stageyellow}{RGB}{255,209,0}

\begin{document}

\title{Learning to Configure: Data-Driven Joint Resource Allocation Optimization in RIS-assisted Networks}

\title{CALO: Constraint-Aware Learning Optimization for Joint Resource Allocation in Double-Active RIS-Assisted Wireless Networks}

\author{Alaa~S.~Arabiyat and Mohammad~J.~Abdel-Rahman,~\IEEEmembership{Senior Member,~IEEE}
\thanks{A.~S.~Arabiyat is with the Department of Data Science, Princess Sumaya University for Technology, Amman 11941, Jordan (e-mail:alaa.arabiyat@hotmail.com ).}
\thanks{M.~J.~Abdel-Rahman is with the Department of Data Science, Princess Sumaya University for Technology, Amman 11941, Jordan. He is also with the Department of Electrical and Computer Engineering, Virginia Tech, Blacksburg, VA 24061, USA (e-mail: m.Abdel-Rahman@psut.edu.jo).}
\thanks{This work was supported by Jordan Scientific Research and Innovation Support Fund (SRISF) under Grant ICT/1/21/2021.}
}

\markboth{IEEE Transactions on Cognitive Communications and Networking}%
{Arabiyat \MakeLowercase{\textit{et al.}}: CALO: Constraint-Aware Learning Optimization for Joint Resource Allocation in Double-Active RIS-Assisted Wireless Networks}


\maketitle

\begin{abstract}
Double-active reconfigurable intelligent surface (RIS)-assisted wireless systems can improve coverage and achievable rate in blockage-dominated environments. Still, their joint resource allocation is challenging due to the coupling among RIS placement, amplification power allocation, and reflecting-element assignment. The resulting problem is linearly constrained, non-convex, and involves both continuous and discrete variables, making conventional iterative solvers such as block coordinate descent (BCD) computationally expensive for real-time deployment. This paper proposes a \underline{c}onstraint-\underline{a}ware \underline{l}earning \underline{o}ptimization (CALO) framework for data-driven joint resource allocation in double-active RIS-assisted networks. CALO reformulates the decision variables into grouped fractional representations and maps them to physical resources through constraint-preserving transformations, ensuring that distance, power, and element-budget constraints are satisfied by construction. A straight-through estimator is incorporated to enable differentiable learning over discrete reflecting-element assignments, while a regret-driven hinge objective uses the BCD solution as a reference and encourages performance improvement beyond solver imitation. Simulation results show that CALO achieves $100\%$ feasibility across all tested configurations, improves the achievable rate over BCD in both urban and rural scenarios, and reduces online inference time by orders of magnitude. These results demonstrate the effectiveness of structure-aware learning for feasible and real-time optimization in active multi-RIS wireless systems.
\end{abstract}



\section{Introduction}

\IEEEPARstart{T}{he} sixth generation (6G) of wireless communication systems is expected to support unprecedented requirements in terms of data rate, latency, energy efficiency, reliability, and network adaptability. These requirements are driven by emerging applications such as digital twins~\cite{Applications_For_6G_DT,Applications_For_6G_DT1}, holographic communications~\cite{Applications_For_6G_HC,Applications_For_6G_HC1}, and ultra-reliable low-latency communication (URLLC)~\cite{Applications_For_6G_URLLC,Applications_For_6G_URLLC1}. To meet these demands, higher frequency bands, including sub-THz and THz bands, have been widely investigated due to their abundant spectral resources. However, communication at these frequencies is highly vulnerable to severe path loss, blockage, and strong line-of-sight (LoS) dependency, which limits coverage and link reliability in practical deployments.

Reconfigurable intelligent surfaces (RISs) have emerged as a promising technology to address these propagation challenges by enabling programmable control of the wireless environment. By adjusting the phase and amplitude response of incident electromagnetic waves, RISs can enhance signal strength, improve coverage, and increase spectral and energy efficiency~\cite{RIS_signal_strength,RIS_Book_Chapter}. While early RIS studies mainly considered passive single-surface deployments, recent research has moved toward more advanced architectures, including active RISs, multi-RIS systems, and double-reflection links. These architectures provide additional degrees of freedom for controlling signal propagation and can offer substantial performance gains over single-RIS systems~\cite{multiple_vs_single_1,multiple_vs_single_2,multiple_vs_single_3}.

Among these architectures, double-active RIS systems are particularly attractive in blockage-dominated environments, where the direct transmitter-receiver link and single-reflection paths may be unavailable. By employing two active surfaces, the system can exploit a cascaded double-reflection link while compensating for severe attenuation through signal amplification. However, these benefits introduce a more challenging resource-allocation problem. The achievable rate depends jointly on RIS placement, amplification power allocation, and reflecting-element assignment. These variables are strongly coupled through the cascaded channel and are subject to strict physical constraints, including total distance, total amplification power, and total number of reflecting elements.

The resulting optimization problem is non-convex, mixed continuous-discrete, and linearly constrained. Classical optimization methods, such as alternating optimization (AO) and block coordinate descent (BCD), address this problem by decomposing it into tractable subproblems and iteratively updating the decision variables~\cite{Liu2024Survey,Razaviyayn2020SPM}. Although such methods can provide high-quality reference solutions, they require repeated optimization for every system configuration. Their computational complexity increases with the number of RIS elements, system parameters, and optimization blocks, making them difficult to deploy in real-time or dynamic wireless environments.

Learning-based optimization has recently been explored as an alternative to iterative wireless resource allocation. Neural networks can approximate the mapping from system parameters to resource-allocation decisions and provide low-latency inference after offline training. However, existing learning-based approaches still face two major limitations. First, many methods treat resource allocation as a black-box prediction problem and do not guarantee strict feasibility with respect to equality, budget, or integer constraints. Second, supervised learning approaches often imitate solutions generated by classical solvers, which may cause the learned model to inherit the suboptimality of the reference algorithm rather than improve upon it.

To address these limitations, this paper proposes a \emph{\underline{c}onstraint-\underline{a}ware \underline{l}earning \underline{o}ptimization} (CALO) framework for joint resource allocation in double-active RIS-assisted wireless systems. Instead of directly predicting unconstrained physical variables, CALO reformulates the problem using fractional grouped decision variables. These variables are mapped to physical resources through constraint-preserving transformations, ensuring that the placement, power-allocation, and element-assignment constraints are satisfied by construction. Discrete reflecting-element allocation is handled through a straight-through estimator (STE), enabling end-to-end training despite the presence of integer-valued variables. Moreover, a regret-driven hinge objective is introduced to use BCD as a performance reference while encouraging the learned solution to exceed the baseline achievable rate.

From a broader perspective, the proposed framework is not limited to a single double-active RIS configuration. Rather, it represents a learning-based solver for a class of linearly constrained non-convex wireless resource-allocation problems with grouped continuous and discrete variables. By embedding the structure of the optimization problem into the neural architecture, CALO combines the efficiency of single-shot inference with strict feasibility guarantees and direct performance optimization.

\textbf{Our Contributions:}
After formulating the joint optimization of RIS placement, amplification power allocation, and reflecting-element assignment in double-active RIS-assisted wireless systems as a grouped, linearly constrained, non-convex optimization problem with mixed continuous and discrete variables, our main contributions of this paper can be summarized as follows:

\begin{itemize}
\item \textbf{Feasibility-by-construction neural parameterization:} We develop a fractional resource-allocation parameterization based on grouped normalization, where the predicted variables are mapped to physical resources while satisfying the distance, power, and element-budget constraints by construction.
\item \textbf{End-to-end handling of discrete element allocation:} We incorporate an STE to enable differentiable learning over integer reflecting-element assignments, allowing continuous and discrete decision variables to be optimized within a unified neural framework.
\item \textbf{Regret-driven learning beyond solver imitation:} We introduce a hinge-style regret objective that uses BCD-generated solutions as reference points while penalizing the model only when its achievable rate falls below the baseline, thereby encouraging performance improvement rather than direct imitation.
\end{itemize}

We show through extensive simulations that the proposed CALO framework achieves strictly feasible solutions, statistically significant achievable-rate gains over BCD, and orders-of-magnitude reduction in online inference time across heterogeneous deployment scenarios.

\textbf{Paper Organization:}
The remainder of this paper is organized as follows. Section~\ref{sec:RelatedWork} reviews the related literature. Section~\ref{sec:SystemModel} presents the system model and problem formulation. Section~\ref{sec:Method} details the proposed CALO framework. Section~\ref{sec:Results} reports the simulation results and performance analysis. Finally, Section~\ref{sec:Conclusion} concludes 
the paper.

\section{Related Work}\label{sec:RelatedWork}

RISs have emerged as a key technology for beyond-5G and 6G networks by enabling programmable control of the wireless propagation environment. Early studies mainly considered passive single-RIS architectures, where transmit beamforming and RIS phase-shift design are the main optimization variables. However, passive RISs suffer from severe multiplicative path loss in cascaded transmitter-RIS-receiver links, especially in high-frequency, long-distance, and blockage-dominated scenarios. To mitigate this limitation, active RIS architectures have been proposed, where RIS elements provide signal amplification in addition to phase control. While active RISs can compensate for propagation loss, they introduce amplification noise, hardware complexity, and power-budget constraints~\cite{ahmed2025activeRISsurvey,gavriilidis2025activeRIScircuit}. Therefore, power allocation becomes a key design dimension in active RIS systems~\cite{zhi2022activePassivePower}. More recently, multi-RIS and cooperative RIS architectures have been investigated to exploit additional reflection paths and spatial degrees of freedom in complex environments~\cite{pan2025wmmseMultiRIS,xue2026multiIRSmmWave}. In double-active RIS systems, this coupling becomes more pronounced because two active surfaces jointly determine the cascaded signal path, amplification noise, power distribution, and reflecting-element allocation. Existing works have studied deployment optimization, active element allocation, and total power-constrained designs~\cite{kang2023doubleActiveIRS,li2023doubleActiveIRS}, providing the main architectural foundation for the system considered in this paper.

Resource allocation in RIS-assisted systems is generally non-convex due to the coupling among beamforming, RIS coefficients, deployment variables, power allocation, user association, and hardware constraints. This challenge is further amplified in active and multi-RIS systems, where power budgets, noise enhancement, multiple reflecting surfaces, and distributed channels introduce additional coupled variables. Conventional optimization methods, including AO, BCD, weighted minimum mean-square error (WMMSE), successive convex approximation, semidefinite relaxation, and manifold optimization, remain dominant for solving such non-convex wireless design problems~\cite{liu2024optimizationSurvey,ahmed2025RISresourceSurvey}. For example, WMMSE-based designs have been applied to multi-RIS-assisted cell-free networks~\cite{pan2025wmmseMultiRIS}, while robust and secure STAR-RIS systems commonly rely on iterative optimization to handle imperfect CSI, secrecy constraints, and coupled beamforming variables~\cite{pala2025starRISsecurity}. Although these methods provide strong reference solutions, they require repeated per-instance optimization, which limits scalability and real-time applicability, particularly in active and double-active RIS systems~\cite{kang2023doubleActiveIRS,li2023doubleActiveIRS}.

Learning-based optimization has recently emerged as an alternative for reducing online computational complexity in wireless resource allocation. Neural models can learn mappings from system parameters to resource-allocation decisions, enabling fast inference after offline training. Recent surveys highlight the growing role of supervised learning, reinforcement learning, deep reinforcement learning, and graph-based learning in wireless resource management~\cite{pivoto2025mlResourceAllocationSurvey,dai2025graphLearningResourceManagement}. In RIS-assisted systems, learning-based methods have been applied to phase-shift design, beamforming, channel estimation, and resource allocation. However, many existing approaches follow black-box prediction or policy-learning paradigms and do not explicitly guarantee feasibility with respect to equality, power-budget, integer, or hardware constraints. Moreover, supervised models trained only to imitate AO/BCD solutions may inherit the suboptimality of the reference solver. Recent works comparing optimization and machine-learning solutions for advanced RIS architectures highlight the promise of learning-based methods, but also reveal the need for stronger integration of problem structure into the learning model~\cite{pala2025starRISsecurity,liu2025dynamicL2O}.

In contrast, the proposed CALO framework embeds the linear resource-allocation constraints directly into the neural parameterization. It maps grouped fractional variables through simplex-preserving transformations to satisfy placement, power-allocation, and reflecting-element budget constraints by construction, handles discrete element allocation using a straight-through estimator, and adopts a regret-driven hinge objective that encourages performance improvement beyond BCD imitation. Hence, CALO is tailored to linearly constrained non-convex RIS resource-allocation problems involving both continuous and discrete variables.

\section{System Model and Problem Formulation}\label{sec:SystemModel}

\subsection{General Problem Structure and Motivation}

The framework developed in this paper targets a class of wireless resource allocation problems that can be expressed as structured linearly constrained non-convex optimization programs with coupled continuous and discrete decision variables. A generic form of this class can be written as:
\begin{align}
\max_{\mathbf{z}} \quad & f(\mathbf{z};\mathbf{u}) \label{eq:generic_problem}\\
\text{s.t.} \quad
& \sum_{j \in \mathcal{G}_m} z_j = B_m,\quad \forall m \in \mathcal{M}, \label{eq:generic_budget}\\
& z_j \ge 0,\quad \forall j, \label{eq:generic_nonnegative}\\
& z_k \in \mathbb{Z},\quad \forall k \in \mathcal{I}, \label{eq:generic_integer}
\end{align}
where $\mathbf{z}$ denotes the vector of decision variables, $\mathbf{u}$ collects the system parameters, $\{\mathcal{G}_m\}_{m \in \mathcal{M}}$ represents groups of variables subject to linear budget constraints, and $\mathcal{I}$ denotes the subset of integer-valued variables.

The objective function $f(\mathbf{z};\mathbf{u})$ is generally non-convex due to multiplicative coupling among variables, cascaded signal transformations, and nonlinear dependence on system parameters. The constraints in \eqref{eq:generic_budget} define a set of affine budget constraints that induce a structured feasible region, where decision variables are organized into groups sharing limited resources such as power, distance, or hardware elements.

This class of problems frequently arises in advanced wireless systems, where multiple interdependent resources must be jointly optimized under physical constraints. Their non-convexity, variable coupling, and mixed decision spaces make conventional optimization methods computationally demanding, especially in large-scale or real-time settings.

The double-active RIS-assisted system considered in this paper is a representative instance of the problem class defined in \eqref{eq:generic_problem}--\eqref{eq:generic_integer}. The joint optimization of RIS placement, amplification power allocation, and reflecting-element assignment gives rise to grouped variables under linear resource constraints and a strongly coupled non-convex objective induced by multi-stage signal propagation and amplification. This formulation exposes a structured optimization template that facilitates feasibility-preserving parameterizations and scalable learning-based inference, allowing the proposed framework to serve as a general solver for a broader class of linearly constrained non-convex optimization problems.

\subsection{System Model}

\begin{figure}[]
\centering
\includegraphics[width=2.5in]{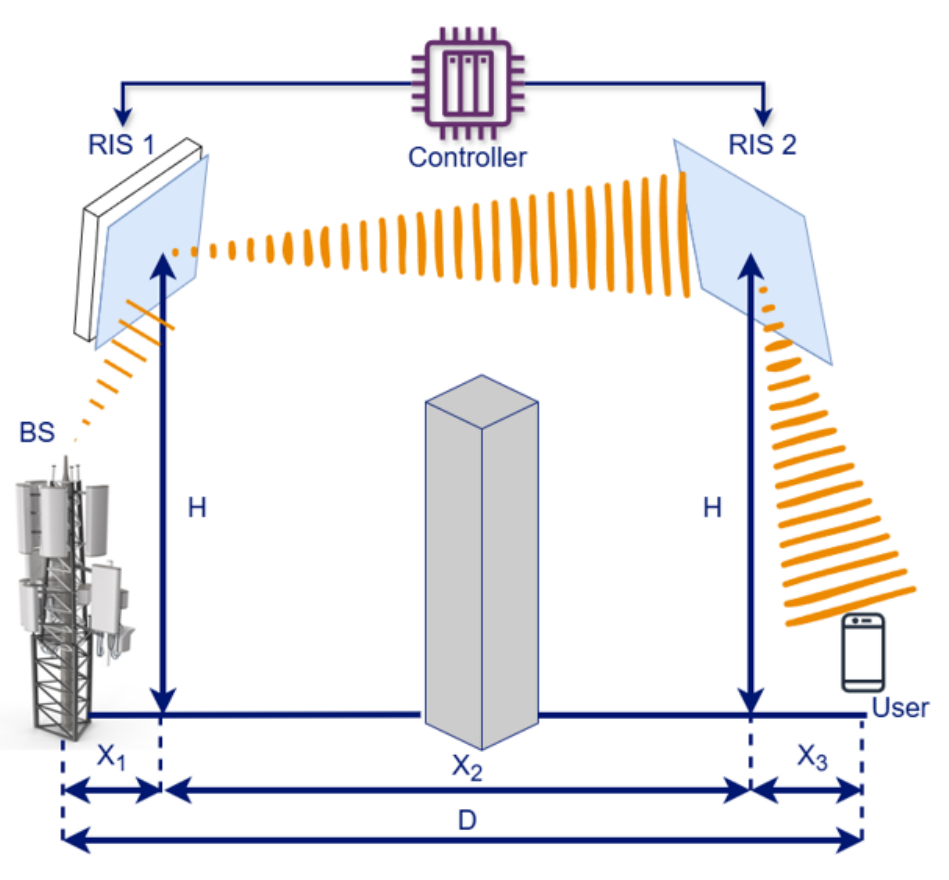}
\caption{Double-active RIS-assisted wireless communication system.}
\label{fig:system_model}
\end{figure}

We consider a downlink wireless communication system assisted by two cooperative active RISs, as illustrated in Fig.~\ref{fig:system_model}. A base station (BS) equipped with $M$ antennas communicates with a single-antenna user through a double-reflection link formed by RIS~$1$ and RIS~$2$. This system setup follows the model in~\cite{Ref15}, while being adopted here as a representative structure to expose the underlying optimization challenges.
\subsubsection{Network Geometry}

The BS, RIS~$1$, RIS~$2$, and the user are deployed along a horizontal line. Let $x_1$, $x_2$, and $x_3$ denote the horizontal distances of the $\text{BS} \rightarrow \text{RIS}~1$, $\text{RIS}~1 \rightarrow \text{RIS}~2$, and $\text{RIS}~2 \rightarrow \text{user}$ links, respectively, satisfying $x_1 + x_2 + x_3 = D$, where $D$ is the total BS-user distance. Both RISs are deployed at a fixed height $H$. The total number of reflecting elements is $N$, which is partitioned as $N_1 + N_2 = N, N_1, N_2 \in \mathbb{N}$. We focus on a challenging scenario where the direct BS-user link and single-reflection paths are blocked, and only the cascaded $\text{BS} \rightarrow \text{RIS}~1 \rightarrow \text{RIS}~2 \rightarrow \text{user}$ link is available, as commonly considered in multi-RIS systems~\cite{Ref15}.

\subsubsection{Channel Model}


To provide clear structural insights, we assume that all channels have a direct LoS, following~\cite{Ref15}. The BS-to-RIS~$1$ channel is $\mathbf{H}_B = h_B \mathbf{a}_r \mathbf{a}_t^H$, where $h_B = \frac{\sqrt{\beta}}{d_1} e^{-j\frac{2\pi}{\lambda} d_1}$ represents the path-loss and phase shift, with $d_1 = \sqrt{x_1^2 + H^2}$. Similarly, the inter-RIS channel $\mathbf{H}_I \in \mathbb{C}^{N_2 \times N_1}$ and the RIS~$2$-to-user channel $\mathbf{h}_U^H \in \mathbb{C}^{1 \times N_2}$ are modeled using the same LoS structure.

\subsubsection{Active RIS Model}

Each RIS performs both reflection and amplification. The reflection matrix of RIS $i \in \{1,2\}$ is $\boldsymbol{\Theta}_i = \mathrm{diag}\left(a_i e^{j\phi_{i,1}}, \ldots, a_i e^{j\phi_{i,N_i}}\right)$, where $a_i$ is the amplification factor and $\phi_{i,n}$ is the phase shift of the $n$-th element.

Due to hardware constraints and under the LoS assumption, all elements within each RIS share a common amplification factor, as justified in~\cite{Ref15}. The amplification process introduces noise $\mathbf{n}_i \sim \mathcal{CN}(0, \delta_i^2 \mathbf{I}), i \in \{1,2\}$ at RIS~$i$, while the receiver noise is $n_0 \sim \mathcal{CN}(0, \delta_0^2)$. The BS applies a beamforming vector $\mathbf{w} \in \mathbb{C}^{M}$ satisfying $\|\mathbf{w}\|^2 \le 1$.

\subsubsection{Received Signal Model}

The received signal at the user is expressed as:
\begin{equation}
y = \mathbf{h}_U^H a_2 \boldsymbol{\Theta}_2 
\left(
\mathbf{H}_I a_1 \boldsymbol{\Theta}_1 \mathbf{H}_B \mathbf{w} s
+ \mathbf{H}_I a_1 \boldsymbol{\Theta}_1 \mathbf{n}_1
+ \mathbf{n}_2
\right) + n_0, \nonumber
\end{equation}
where $s$ denotes the transmitted symbol with power $P_B$. This expression captures the cascaded signal propagation and the accumulation of amplification noise across RIS stages.

\subsubsection{Achievable Rate}

The signal-to-noise ratio (SNR) at the user can be written as:
\begin{equation}
\gamma =
\frac{
P_B \left| \mathbf{h}_U^H a_2 \boldsymbol{\Theta}_2 \mathbf{H}_I a_1 \boldsymbol{\Theta}_1 \mathbf{H}_B \mathbf{w} \right|^2
}{
\delta_1^2 \|\mathbf{h}_U^H a_2 \boldsymbol{\Theta}_2 \mathbf{H}_I a_1 \boldsymbol{\Theta}_1 ||^2 + \delta_2^2 \|\mathbf{h}_U^H a_2 \boldsymbol{\Theta}_2 ||^2 + \delta_0^2
}. \nonumber 
\end{equation}

The corresponding achievable rate is $R = \log_2(1 + \gamma)$.

\subsubsection{Remarks on Model Generality}

Although the above model follows a LoS assumption for analytical tractability, the formulation captures the essential characteristics of multi-stage RIS-assisted communication systems. The structure can be extended to more general channel models without altering the underlying optimization framework.

\subsection{Problem Formulation}

Let $P_1$ and $P_2$ denote the amplification power allocated to RIS~$1$ and RIS~$2$, respectively, subject to a total amplification power budget. The objective is to maximize the achievable downlink rate by jointly optimizing the BS beamforming vector, RIS configurations, amplification factors, element allocation, power allocation, and placement variables. Accordingly, the joint optimization problem is formulated as:
\begin{align}
\text{(P1)} \quad
& \max_{\mathbf{w}, \boldsymbol{\Theta}, \mathbf{a}, \mathbf{N}, \mathbf{P}, \mathbf{X}} \quad
 \log_2(1 + \gamma) \\
\text{s.t.} \quad
& \|\mathbf{w}\|^2 \le 1, \\
& \boldsymbol{\Theta}_i =
\mathrm{diag}\big(a_i e^{j\phi_{i,1}}, \ldots, a_i e^{j\phi_{i,N_i}}\big), \quad i \in \{1,2\},
\\
& N_1 + N_2 = N,\quad N_1, N_2 \in \mathbb{N}, \\
& P_1 + P_2 = P_F,\quad P_1, P_2 \ge 0, \\
& x_1 + x_2 + x_3 = D,\quad x_i \ge 0, \\
& a_1^2 \mathcal{P}_1(\cdot) \le P_1, \\
& a_2^2 \mathcal{P}_2(\cdot) \le P_2.
\end{align}
where $\boldsymbol{\Theta} = \{\boldsymbol{\Theta}_1, \boldsymbol{\Theta}_2\}$, $\mathbf{a} = \{a_1, a_2\}$, $\mathbf{N} = \{N_1, N_2\}$, $\mathbf{P} = \{P_1, P_2\}$, and $\mathbf{X} = \{x_1, x_2, x_3\}$. The functions $\mathcal{P}_1(\cdot)$ and $\mathcal{P}_2(\cdot)$ represent the received signal power at RIS~$1$ and RIS~$2$, respectively, and are directly derived from the signal model. These constraints ensure that the amplification power at each RIS does not exceed its allocated budget.


Problem P1 is highly challenging due to several intrinsic properties:

\begin{itemize}
\item \textbf{Non-convex objective:} The achievable rate depends on the SNR, which involves multiplicative coupling across beamforming, RIS configurations, amplification, and placement variables.

\item \textbf{Mixed discrete-continuous variables:} The element allocation variables $\{N_1, N_2\}$ are integer-valued, while the remaining variables are continuous.

\item \textbf{Strong variable coupling:} All decision variables jointly affect both signal power and noise propagation due to the cascaded structure of the system.

\item \textbf{Coupled constraints:} The amplification constraints depend on multiple variables simultaneously, making the feasible region highly non-trivial.
\end{itemize}

These characteristics render P1 a linearly constrained non-convex optimization problem with mixed variables.
Although P1 involves multiple coupled decision variables, it admits a structured decomposition that significantly reduces its effective dimensionality. In particular, following the approach in~\cite{Ref15}, it can be shown that for any given element allocation, power allocation, and placement, the optimal BS beamforming vector, RIS phase shifts, and amplification factors can be obtained in closed form.




\subsection{Problem Reformulation and Structural Reduction}


\subsubsection{Closed-Form Elimination of Auxiliary Variables}

Following \cite{Ref15}, for any feasible element allocation $\mathbf{N}$, amplification power allocation $\mathbf{P}$, and placement $\mathbf{X}$, the auxiliary variables, namely the BS beamforming vector $\mathbf{w}$, RIS phase shifts $\boldsymbol{\Theta}_1, \boldsymbol{\Theta}_2$, and amplification factors $a_1, a_2$,admit closed-form optimal solutions.
\begin{align}
    \mathbf{w}^* &= \frac{\boldsymbol{\alpha}_t(\theta_B^t, M)}{\|\boldsymbol{\alpha}_t(\theta_B^t, M)\|}, \nonumber \\[6pt]
    \big[\boldsymbol{\Theta}_1^*\big]_{n_1} &= 
    e^{-j\left(\angle \big[\boldsymbol{\alpha}_t^H(\theta_I^t, \vartheta_I^t, N_1)\big]_{n_1}
    + \angle \big[\boldsymbol{\alpha}_r(\theta_B^r, \vartheta_B^r, N_1)\big]_{n_1}\right)}, \nonumber \\[6pt]
    \big[\boldsymbol{\Theta}_2^*\big]_{n_2} &= 
    e^{-j\left(\angle \big[h_U^H\big]_{n_2}
    + \angle \big[\boldsymbol{\alpha}_r(\theta_I^r, \vartheta_I^r, N_2)\big]_{n_2}\right)}, \nonumber \\[6pt]
    a_1^* &= \sqrt{\frac{P_1}{N_1 (M P_B \eta_B^2 + \delta_1^2)}}, \nonumber \\[6pt]
    a_2^* &= \sqrt{\frac{P_2 (\delta_1^2 + M P_B \eta_B^2)}
    {N_2 \left(\delta_1^2 (\delta_2^2 + P_1 \eta_I^2) + M P_B \eta_B^2 (\delta_2^2 + N_1 P_1 \eta_I^2)\right)}}, \nonumber 
\end{align}

where
\[
\eta_B = \frac{\sqrt{\beta}}{d_1}, \quad
\eta_I = \frac{\sqrt{\beta}}{d_2}, \quad
\eta_U = \frac{\sqrt{\beta}}{d_3}.
\]

These results imply that $\mathbf{w}$, $\boldsymbol{\Theta}_1$, $\boldsymbol{\Theta}_2$, $a_1$, and $a_2$ can be treated as auxiliary variables and eliminated via substitution.

\subsubsection{Equivalent SNR Expression}

Substituting the optimal auxiliary variables into the received signal model yields an equivalent SNR expression:
\begin{equation}
\gamma =
\frac{N_1^2 N_2^2 M P_B \eta_B^2 \eta_I^2 \eta_U^2}
{N_1 N_2 \delta_1^2 \eta_I^2 \eta_U^2
+ \frac{N_2 \delta_2^2 \eta_U^2}{a_1^2}
+ \frac{\delta_0^2}{a_1^2 a_2^2}}. \nonumber 
\label{eq:gamma_main}
\end{equation}

Further substituting $a_1^*$ and $a_2^*$ leads to a reduced-form expression:
\begin{equation}
\gamma =
\frac{M P_B \beta^3}
{\frac{E_1}{N_1 N_2}
+ \frac{E_2}{N_2 P_2}
+ \frac{E_3}{N_1 N_2 P_1}
+ \frac{E_4}{N_1 N_2 P_1}
+ \frac{E_5}{N_1 N_2 P_2}
+ \frac{E_6}{N_1 N_2 P_1 P_2}}, \nonumber
\label{eq:gamma_reduced}
\end{equation}

where $E_1, \dots, E_6$ are constants determined by system parameters (path loss, transmit power, and noise levels).

\subsubsection{Reduced Problem Formulation}

Therefore, P1 is equivalently reduced to:
\begin{align}
\text{(P2)} \quad
\max_{\mathbf{N}, \mathbf{P}, \mathbf{X}} \quad
& \gamma(\mathbf{N}, \mathbf{P}, \mathbf{X}) \\
\text{s.t.} \quad
& N_1 + N_2 = N,\quad N_1, N_2 \in \mathbb{N}, \label{eqn:P2N}\\
& P_1 + P_2 = P_F,\quad P_1, P_2 \ge 0, \label{eqn:P2P}\\
& x_1 + x_2 + x_3 = D,\quad x_i \ge 0. \label{eqn:P2x}
\end{align}

This reformulation significantly reduces the dimensionality of the problem while preserving its non-convex nature.

\subsubsection{Structural Interpretation}

The reduced problem reveals a grouped structure where decision variables are partitioned into three independent resource allocation groups: Element allocation $\{N_1, N_2\}$, power allocation $\{P_1, P_2\}$, and placement $\{x_1, x_2, x_3\}$. Each group is constrained by a linear equality (budget constraint), forming a simplex domain. This structure aligns with the generic formulation introduced in Section II-A.

\subsubsection{Limitations of Conventional Optimization}

The reference solution applies BCD to iteratively optimize each variable group. However, such approaches suffer from high computational complexity, limited scalability, and a lack of real-time applicability in dynamic wireless environments.

\subsubsection{Transition to Learning-Based Reformulation}

To overcome these limitations, we reformulate the decision variables using fractional representations with simplex constraints.
This transformation guarantees feasibility by construction and converts P2 into an unconstrained optimization problem over simplex domains. This reformulation enables the design of a learning-based framework that directly predicts optimal resource allocation in a single forward pass.

\section{CALO Framework}\label{sec:Method}


\subsection{Framework Overview}

\begin{figure}[!t]
\centering
\resizebox{0.42\textwidth}{!}{
\begin{tikzpicture}[
>=Latex,
font=\footnotesize,
stagebox/.style={
    draw=black!85,
    rounded corners=8pt,
    line width=1.1pt,
    inner sep=2pt
},
box/.style={
    rectangle,
    rounded corners=6pt,
    draw=black!75,
    line width=0.7pt,
    align=center,
    minimum height=1.1cm,
    text width=1.8cm
},
softbox/.style={
    rectangle,
    rounded corners=6pt,
    draw=black!75,
    line width=0.7pt,
    align=center,
    minimum height=1.1cm,
    text width=1.9cm,
    fill=stagepeach
},
arrow/.style={-Latex, line width=0.9pt},
backarrow/.style={-Latex, dashed, line width=1pt}
]

\node[box,fill=stagegreen] (input) at (0,0)
{Input\\Features};

\node[box,fill=stageblue] (mlp) at (2.7,0)
{MLP\\Predictor};

\node[softbox] (g2) at (5.4,0)
{Grouped\\Softmax\\Mapping};

\node[box,fill=stagelilac] (map) at (5.4,-3.2)
{Physical\\Variable\\Mapping};

\node[box,fill=stagelilac] (ste) at (2.7,-3.2)
{STE for\\Discrete\\Variable};

\node[box,fill=stageyellow] (obj) at (0,-3.2)
{Compute\\Objective};

\node[box,fill=stageyellow] (loss) at (0,-4.8)
{Margin-Based\\Loss};

\draw[arrow] (input) -- (mlp);
\draw[arrow] (mlp) -- (g2);
\draw[arrow] (g2) -- (map);
\draw[arrow] (map) -- (ste);
\draw[arrow] (ste) -- (obj);
\draw[arrow] (obj) -- (loss);

\coordinate (bp1) at (-1.6,-4.8);
\coordinate (bp2) at (-1.6,-1.5);

\draw[backarrow]
(loss.west) -- (bp1)
node[midway,above=3pt,align=center]
{}
-- (bp2)
-| (mlp.south);

\node[align=center,font=\footnotesize]
at (0.5,-1.5)
{\textbf{Backpropagation}\\Update neural network weights};

\node[stagebox,fit=(input)] (s1) {};
\node[stagebox,fit=(mlp)] (s2) {};
\node[stagebox,fit=(g2)] (s3) {};
\node[stagebox,fit=(map)(ste)] (s4) {};
\node[stagebox,fit=(obj)(loss)] (s5) {};

\node[above=1pt of s1,font=\bfseries] {Stage 1};

\node[above=1pt of s2,font=\bfseries] {Stage 2};

\node[above=1pt of s3,font=\bfseries] {Stage 3};

\node[above=1pt of s4,font=\bfseries] {Stage 4};

\node[above=1pt of s5,font=\bfseries] {Stage 5};

\end{tikzpicture}
}
\caption{Overall proposed learning framework. The neural network predicts grouped fractional allocations, which are transformed through softmax-based heads, mapped into physical variables, and refined using an STE for discrete variables. The objective and margin-based loss are then computed, and the network parameters are updated via backpropagation.}
\label{fig:framework_overall}
\end{figure}
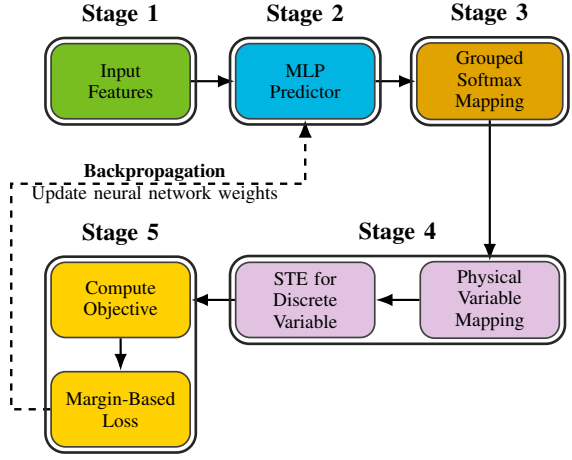


To solve the reduced structured non-convex optimization problem (P2), we propose the \textit{constraint-aware learning optimization} (CALO) framework, which replaces iterative optimization with a direct learning-based solution. Specifically, CALO learns a mapping:
\begin{equation}
\mathcal{F}_\theta: \mathbf{u} \rightarrow (\mathbf{X}, \mathbf{P}, \mathbf{N}),
\end{equation}
where $\mathbf{u}$ denotes the system parameters, including $D$, $N$, $P_F$, and channel-related variables. The mapping is parameterized by a neural network and produces a high-quality feasible solution in a single forward pass.


As illustrated in Fig.~\ref{fig:framework_overall}, the framework consists of a neural predictor that outputs unconstrained logits, which are transformed via grouped softmax operations into fractional variables defined over simplex domains. These structured transformations enforce feasibility by construction and map the fractional variables into the corresponding physical decision variables. Discrete variables are handled using an STE, enabling end-to-end differentiable training. The resulting variables are then evaluated through the system model to compute the achievable rate, which is used to define a performance-driven loss.

Unlike conventional black-box learning approaches, CALO explicitly incorporates the structure of the underlying optimization problem into the model design. This ensures feasibility, enables direct optimization of the system performance, and avoids reliance on iterative solvers. As a result, CALO replaces computationally expensive iterative optimization with a single-shot inference mechanism, enabling efficient and real-time resource allocation in RIS-assisted wireless systems.

\subsection{Challenges in Learning-Based Optimization}

Despite their potential, the application of learning-based methods to structured resource allocation problems, such as P2, faces several fundamental challenges.

\textbf{Constraint awareness:} Deep neural networks are unconstrained function approximators and do not inherently satisfy problem-specific feasibility conditions. In the considered problem, the decision variables are subject to strict linear equality constraints. Conventional approaches rely on penalty terms or projection steps, which may lead to infeasible or suboptimal solutions.

\textbf{Absence of optimal labels:} The underlying problem is non-convex, and globally optimal solutions are generally intractable. Existing methods such as BCD provide only suboptimal solutions, limiting the applicability of supervised learning approaches that depend on high-quality ground truth labels.

\textbf{Discrete decision variables:} The presence of discrete variables, such as reflecting-element allocation, introduces non-differentiability and prevents direct application of gradient-based optimization. This challenge is further exacerbated by the coupling between discrete and continuous variables.

These challenges motivate the design of the proposed CALO framework, which addresses them through constraint-aware parameterization, differentiable handling of discrete variables, and performance-driven learning.

\subsection{Constraint-Aware Parameterization}

A central challenge in P2 is ensuring that the predicted solutions satisfy the linear equality constraints~\eqref{eqn:P2N}-\eqref{eqn:P2x}
together with non-negativity and integer requirements. To address this, CALO adopts a structure-aware parameterization that transforms the constrained variables into fractional representations defined over simplex domains.

\subsubsection{Fractional Representation}

We introduce three groups of fractional variables: $\boldsymbol{\alpha} = [\alpha_1, \alpha_2, \alpha_3]$, $\boldsymbol{\beta} = [\beta_1, \beta_2]$, and $\boldsymbol{\gamma} = [\gamma_1, \gamma_2]$, such that:
\begin{align}
\sum_{i=1}^{3} \alpha_i &= 1, \quad \alpha_i \ge 0, \label{eq:simplex_alpha} \\
\sum_{i=1}^{2} \beta_i &= 1, \quad \beta_i \ge 0, \label{eq:simplex_beta} \\
\sum_{i=1}^{2} \gamma_i &= 1, \quad \gamma_i \ge 0. \label{eq:simplex_gamma}
\end{align}

The physical variables are obtained through:
\begin{align}
x_i &= \alpha_i \; D, \quad i = 1, 2, 3, \label{eq:mapping_x} \\
P_i &= \beta_i \; P_F, \quad i = 1, 2, \label{eq:mapping_P} \\
N_i &= \mathrm{round}(\gamma_i \; N), \quad i = 1, 2. \label{eq:mapping_N}
\end{align}

\subsubsection{Feasibility Guarantee}

By construction, the simplex constraints in~\eqref{eq:simplex_alpha} and~\eqref{eq:simplex_beta}, together with the mappings in~\eqref{eq:mapping_x} and~\eqref{eq:mapping_P}, ensure that the equality constraints in~\eqref{eqn:P2x} and~\eqref{eqn:P2P} are strictly satisfied. For the discrete variables, the rounding operation in \eqref{eq:mapping_N} enforces integer feasibility. Any residual mismatch is resolved through a simple correction step, ensuring that \eqref{eqn:P2N} holds exactly.

\subsubsection{Neural Parameterization via Softmax}

To enforce the simplex constraints, the fractional variables are obtained via softmax transformations applied to the neural network outputs:
\begin{align}
\alpha_i &= \frac{e^{z_{\alpha,i}}}{\sum_{k} e^{z_{\alpha,k}}}, \quad
\beta_i  &= \frac{e^{z_{\beta,i}}}{\sum_{k} e^{z_{\beta,k}}}, \quad
\gamma_i &= \frac{e^{z_{\gamma,i}}}{\sum_{k} e^{z_{\gamma,k}}}. \label{eq:softmax_gamma}
\end{align}

\subsubsection{Discussion}

The proposed parameterization transforms P2 into an unconstrained learning problem over simplex domains, eliminating the need for projection or penalty-based methods while guaranteeing feasibility by construction. Moreover, preserving the grouped structure enables the model to effectively exploit the problem decomposition, leading to accurate and efficient learning.

\subsection{Discrete Variable Handling via STE}

A key challenge in (P2) is the presence of discrete decision variables, namely the reflecting-element allocation $\{N_1, N_2\}$, which must take integer values. This introduces non-differentiability and prevents direct application of gradient-based optimization.

\subsubsection{Forward Mapping}

Using the fractional parameterization in \eqref{eq:mapping_N}, the continuous representation of the discrete variables is obtained as $\tilde{N}_i = \gamma_i N, i = 1,2,$
followed by a rounding operation $N_i = \mathrm{round}(\tilde{N}_i).$

\subsubsection{STE}

To enable gradient-based training, the rounding operation is approximated using STE, where $\frac{\partial N_i}{\partial \tilde{N}_i} \approx 1,$
such that $\frac{\partial \mathcal{L}}{\partial \tilde{N}_i} = \frac{\partial \mathcal{L}}{\partial N_i}.$
This allows gradients to propagate through the discrete variables, enabling end-to-end learning.

\subsubsection{Constraint Consistency}

Since the fractional variables satisfy \eqref{eq:simplex_gamma}, the continuous variables obey $\tilde{N}_1 + \tilde{N}_2 = N.$
After rounding, a small mismatch may occur. To enforce exact feasibility, a simple correction is applied: $N_2 = N - N_1,$
ensuring that \eqref{eqn:P2N} holds exactly.

\subsubsection{Discussion}

The proposed approach enables differentiable learning over discrete variables without resorting to combinatorial search or relaxation techniques. Although STE provides a biased gradient approximation, it has been shown to be effective in practice for learning problems involving discrete operations. In this work, it allows joint optimization of continuous and discrete variables within a unified end-to-end framework while preserving feasibility.

\subsection{Regret-Based Learning Objective}

A key challenge in P2 is the absence of globally optimal ground truth solutions due to its non-convex nature. Existing methods provide only suboptimal solutions, denoted by $R_{\mathrm{ref}}$, which serve as performance baselines rather than optimal labels. Instead of relying on supervised learning, we adopt a performance-driven objective that directly optimizes the achievable rate. Let $R_{\mathrm{pred}}$ denote the rate obtained from the predicted solution. The regret is defined as $R_{\mathrm{ref}} - R_{\mathrm{pred}}$.
The learning objective is formulated as a hinge-based regret minimization:
\begin{equation}
\mathcal{L}_{\mathrm{reg}} = \max \left( 0, \; R_{\mathrm{ref}} - R_{\mathrm{pred}} \right).
\label{eq:regret_loss}
\end{equation}

This objective penalizes the model only when it underperforms the reference solution and imposes zero loss otherwise. As a result, the model is encouraged to match or exceed the performance of the baseline rather than imitate it. The gradient with respect to the network parameters $\theta$ is given by
\begin{equation}
\nabla_{\theta}\mathcal{L}_{\mathrm{reg}}=
\begin{cases}
-\nabla_{\theta}R_{\mathrm{pred}}, & \text{if } R_{\mathrm{pred}}<R_{\mathrm{ref}},\\
0, & \text{otherwise.}
\end{cases}
\end{equation}

This ensures that updates are applied only when the predicted solution is suboptimal, driving the model toward improved performance. Unlike conventional losses, such as MSE, which force imitation of suboptimal labels, the proposed objective enables the model to discover superior solutions.
\begin{figure*}[!t]
\centering
\resizebox{0.9\textwidth}{!}{%
\begin{tikzpicture}[
    x=1.15cm,
    y=0.9cm,
    >=Latex,
    shorten >=1pt,
    font=\small,
    neuron/.style={circle, draw=black!70, thick, minimum size=7mm, inner sep=0pt},
    input neuron/.style={neuron, fill=myGreen!35},
    hidden neuron/.style={neuron, fill=myBlue!20},
    logits neuron/.style={neuron, fill=myRed!25},
    frac neuron/.style={neuron, fill=myPurple!18},
    phys neuron/.style={circle, draw=black!70, thick, minimum size=9mm, inner sep=0pt, fill=myOrange!20},
    softmax/.style={rectangle, rounded corners, draw=myPurple, thick, fill=myPurple!10, minimum width=1.2cm, minimum height=0.8cm, align=center},
    steblock/.style={rectangle, rounded corners, draw=myRed, thick, fill=myRed!10, minimum width=1.4cm, minimum height=0.9cm, align=center},
    mult/.style={circle, draw=black!70, thick, fill=gray!10, minimum size=7mm},
    calcblock/.style={rectangle, rounded corners, draw=black!70, thick, fill=myYellow, minimum width=2.1cm, minimum height=2.2cm, align=center},
    lossblock/.style={rectangle, rounded corners, draw=black!70, thick, fill=green!10, minimum width=2.8cm, minimum height=2.2cm, align=center},
    annot/.style={align=center},
    stagebox/.style={draw=black!55, dashed, rounded corners, inner sep=6pt}
]

\foreach \i/\y in {1/10,2/9,3/8,4/7,5/6,6/5,7/4,8/3,9/2,10/1}
    \node[input neuron] (I-\i) at (0,\y) {};

\node at (0,0.1) {\Large $\vdots$};
\node[annot, above=4pt of I-1] {Input\\Layer};

\foreach \i/\y in {1/9,2/7.6,3/6.2,4/4.8,5/3.4,6/2}
    \node[hidden neuron] (H1-\i) at (2,\y) {};

\foreach \i/\y in {1/9,2/7.6,3/6.2,4/4.8,5/3.4,6/2}
    \node[hidden neuron] (H2-\i) at (3.5,\y) {};

\node[annot] at (2,11) {Hidden\\Layer 1};
\node[annot] at (3.5,11) {Hidden\\Layer n};


\node[logits neuron] (Z-1) at (5,9) {$z_{x1}$};
\node[logits neuron] (Z-2) at (5,8) {$z_{x2}$};
\node[logits neuron] (Z-3) at (5,7) {$z_{x3}$};

\node[logits neuron] (Z-4) at (5,5.5) {$z_{p1}$};
\node[logits neuron] (Z-5) at (5,4.5) {$z_{p2}$};

\node[logits neuron] (Z-6) at (5,3) {$z_{n1}$};
\node[logits neuron] (Z-7) at (5,2) {$z_{n2}$};

\node[annot] at (5,11) {Output\\Logits};


\node[softmax] (SX) at (7,8) {Softmax};
\node[softmax] (SP) at (7,5) {Softmax};
\node[softmax] (SN) at (7,2.5) {Softmax};


\node[frac neuron] (O-1) at (8.7,9) {$\alpha_{x1}$};
\node[frac neuron] (O-2) at (8.7,8) {$\alpha_{x2}$};
\node[frac neuron] (O-3) at (8.7,7) {$\alpha_{x3}$};

\node[frac neuron] (O-4) at (8.7,5.5) {$\alpha_{p1}$};
\node[frac neuron] (O-5) at (8.7,4.5) {$\alpha_{p2}$};

\node[frac neuron] (O-6) at (8.7,3) {$\gamma_{n1}$};
\node[frac neuron] (O-7) at (8.7,2) {$\gamma_{n2}$};

\node[annot] at (8.7,11) {Fractional\\ Variables};


\node[mult] (MX) at (10.2,8) {$\times$};
\node[mult] (MP) at (10.2,5) {$\times$};
\node[mult] (MN) at (10.2,2.5) {$\times$};

\node[rectangle, draw=black!70, thick, fill=blue!10, minimum width=0.9cm, minimum height=0.6cm] (D) at (10.2,9.3) {$D$};
\node[rectangle, draw=black!70, thick, fill=blue!10, minimum width=0.9cm, minimum height=0.6cm] (PF) at (10.2,6.3) {$P_F$};
\node[rectangle, draw=black!70, thick, fill=blue!10, minimum width=0.9cm, minimum height=0.6cm] (N) at (10.2,3.8) {$N$};

\node[steblock] (STE) at (11.5,2.5) {STE};

\node[phys neuron] (A1) at (12.9,9) {$x_1$};
\node[phys neuron] (A2) at (12.9,8) {$x_2$};
\node[phys neuron] (A3) at (12.9,7) {$x_3$};

\node[phys neuron] (A4) at (12.9,5.5) {$P_1$};
\node[phys neuron] (A5) at (12.9,4.5) {$P_2$};

\node[phys neuron, fill=myRed!18] (A6) at (12.9,3) {$N_1$};
\node[phys neuron, fill=myRed!18] (A7) at (12.9,2) {$N_2$};

\node[annot] at (12.9,11) {Physical\\Variables};

\node[calcblock] (RATE) at (16.3,5.5) {Analytical\\Rate Evaluation\\$R_{\mathrm{pred}}$};

\node[lossblock] (LOSS) at (19.0,5.5) {Hinge Loss\\$\max(0,R_{\mathrm{ref}}-R_{\mathrm{pred}})$};

\node[rectangle, rounded corners, draw=black!70, thick, fill=gray!15, minimum width=1.5cm, minimum height=0.8cm] (REF) at (19.0,8.3) {$R_{\mathrm{ref}}$};


\foreach \i in {1,2,3}
    \foreach \j in {1,2,3,4,5,6}
        \draw[black!45] (H1-\i) -- (H2-\j);
\node at (2.6,4) {\Large $\vdots$};
\foreach \i in {1,2,3}
    \foreach \j in {1,2,3,4,5,6,7}
        \draw[black!45] (H2-\i) -- (Z-\j);
\node at (4,4) {\Large $\vdots$};
\node at (1,5.5) {\Large $\cdots$};

\draw[->, thick] (Z-1) -- (SX);
\draw[->, thick] (Z-2) -- (SX);
\draw[->, thick] (Z-3) -- (SX);

\draw[->, thick] (Z-4) -- (SP);
\draw[->, thick] (Z-5) -- (SP);

\draw[->, thick] (Z-6) -- (SN);
\draw[->, thick] (Z-7) -- (SN);

\draw[->, thick] (SX) -- (O-1);
\draw[->, thick] (SX) -- (O-2);
\draw[->, thick] (SX) -- (O-3);

\draw[->, thick] (SP) -- (O-4);
\draw[->, thick] (SP) -- (O-5);

\draw[->, thick] (SN) -- (O-6);
\draw[->, thick] (SN) -- (O-7);

\draw[->, thick] (O-1) -- (MX);
\draw[->, thick] (O-2) -- (MX);
\draw[->, thick] (O-3) -- (MX);
\draw[->, thick] (D) -- (MX);

\draw[->, thick] (O-4) -- (MP);
\draw[->, thick] (O-5) -- (MP);
\draw[->, thick] (PF) -- (MP);

\draw[->, thick] (O-6) -- (MN);
\draw[->, thick] (O-7) -- (MN);
\draw[->, thick] (N) -- (MN);

\draw[->, thick] (MX) -- (A1);
\draw[->, thick] (MX) -- (A2);
\draw[->, thick] (MX) -- (A3);

\draw[->, thick] (MP) -- (A4);
\draw[->, thick] (MP) -- (A5);

\draw[->, thick] (MN) -- (STE);
\draw[->, thick] (STE) -- (A6);
\draw[->, thick] (STE) -- (A7);

\draw[->, thick] (A1.east) -- ++(0.8,0) |- (RATE.west);
\draw[->, thick] (A2.east) -- ++(0.8,0) |- (RATE.west);
\draw[->, thick] (A3.east) -- ++(0.8,0) |- (RATE.west);
\draw[->, thick] (A4.east) -- ++(0.8,0) |- (RATE.west);
\draw[->, thick] (A5.east) -- ++(0.8,0) |- (RATE.west);
\draw[->, thick] (A6.east) -- ++(0.8,0) |- (RATE.west);
\draw[->, thick] (A7.east) -- ++(0.8,0) |- (RATE.west);

\draw[->, thick] (RATE) -- (LOSS);
\draw[->, thick] (REF) -- (LOSS);

\node[stagebox, fit=(I-1)(I-10)] {};
\node[stagebox, fit=(H1-1)(H1-6)(H2-1)(H2-6)(Z-1)(Z-7), label=above:{\textbf{Neural Predictor}}] {};

\node[stagebox, fit=(SX)(SP)(SN)(O-1)(O-7), label=above:{\textbf{Constraint Satisfaction}}] {};

\node[stagebox, fit=(MX)(MP)(MN)(STE)(A1)(A7), label=above:{\textbf{Variable Recovery}}] {};

\node[stagebox, fit=(RATE)(LOSS)(REF), label=above:{\textbf{Objective Evaluation}}] {};

\node[align=left, anchor=west] at (9.7,1.0) {\footnotesize \textcolor{black!75}{Only $N_1,N_2$ are discrete and require STE.}};

\coordinate (p1) at (17.7,0.4);
\coordinate (p2) at (3.5,0.4);
\draw[->, thick] (17.7,4) -- (p1) -- (p2)
node[midway, below=0pt, align=center]
{\textbf{Backpropagation:} Update Neural Network Weights}
-- (3.5,1.4);

\end{tikzpicture}%
}
\caption{Architecture of the proposed neural framework. The multilayer perceptron produces grouped logits that are transformed by separate softmax heads into normalized fractional allocations. These fractions are mapped into physical decision variables using the system constraints. The discrete reflector-count branch is handled by an STE, after which the predicted achievable rate is evaluated and compared against a reference solution through a hinge-style loss.
}
\label{fig:nn_architecture_ste}
\end{figure*}
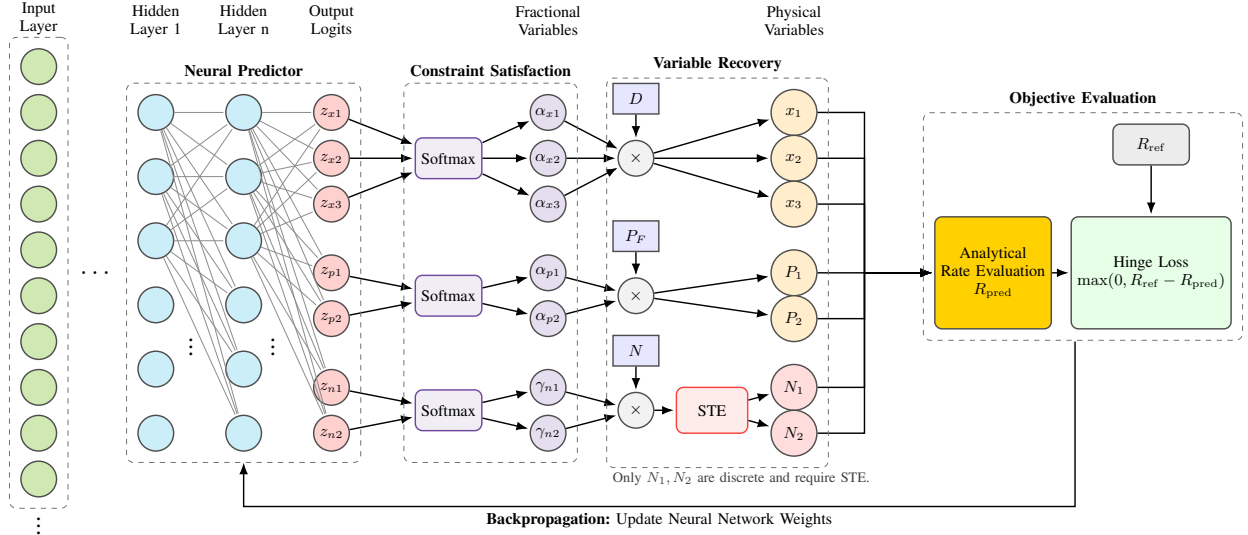

\subsubsection{Discussion}

The proposed formulation shifts the learning objective from solution imitation to direct performance optimization. By leveraging suboptimal solutions as references, the model is encouraged to surpass classical optimization methods while remaining fully compatible with end-to-end gradient-based training.

\subsection{Learning Framework and Model Architecture}

The CALO framework is illustrated in Fig.~\ref{fig:nn_architecture_ste}, where all components are integrated into a unified end-to-end learning pipeline. The model maps system parameters directly to feasible decision variables while optimizing the achievable rate.

\subsubsection{Input and Neural Prediction}

Let $\mathbf{s}$ denote the input vector capturing system parameters, including $D$, $P_F$, and channel-related variables. The input is normalized and fed into a neural network $f_\theta(\cdot)$, which produces grouped logits corresponding to placement, power allocation, and element allocation variables, i.e., $[\mathbf{z}_\alpha, \mathbf{z}_\beta, \mathbf{z}_\gamma] = f_\theta(\mathbf{s})$.

\subsubsection{Structured Variable Construction}

The logits are transformed into fractional variables via softmax mappings as defined in~\eqref{eq:softmax_gamma}. These fractional variables are then mapped to physical decision variables using the deterministic transformations in \eqref{eq:mapping_x}--\eqref{eq:mapping_N}, ensuring feasibility by construction. Discrete variables are handled using the STE mechanism introduced earlier.

\subsubsection{Objective Evaluation}

The resulting decision variables are evaluated through the system model to compute the achievable rate, $R_{\mathrm{pred}} = \mathcal{F}(\mathbf{X}, \mathbf{P}, \mathbf{N}, \mathbf{s})$, 
where $\mathcal{F}(\cdot)$ denotes the analytical rate expression.

\subsubsection{End-to-End Training}

The model is trained using the regret-based loss defined in \eqref{eq:regret_loss}. Gradients propagate through the system model, parameterization layers, and neural network, enabling end-to-end optimization.

\subsubsection{Discussion}

The proposed framework integrates constraint-aware parameterization, discrete variable handling, and performance-driven learning into a single differentiable architecture. This enables direct optimization of the system objective while guaranteeing feasibility and avoiding iterative solvers, resulting in an efficient and scalable solution for RIS-assisted wireless systems.

\section{Simulation Results}\label{sec:Results}

\subsection{Experimental Setup}

The experiments are conducted under a unified and controlled environment to ensure fair comparison across all considered models. A single dataset is generated using BCD, which serves as a strong suboptimal baseline. This ensures consistency in supervision and enables a reliable evaluation of the proposed learning-based framework against the same reference solutions. The dataset is constructed by simulating a wide range of wireless system configurations, capturing variations in key parameters including $D$, $N$, and $P_F$. Such diversity is essential to promote generalization and to evaluate the robustness of the learned models across heterogeneous deployment scenarios.

All experiments are executed on a personal computing platform equipped with an 8-core ARM-based processor (architecture: \texttt{aarch64}) and 8~GB of RAM. Due to the absence of GPU acceleration, both training and inference are performed entirely on CPU resources, highlighting the practicality of the proposed approach under limited computational capabilities. The implementation is developed in Python (version 3.11.6), using TensorFlow (version 2.21.0) as the primary deep learning framework. Supporting libraries include NumPy (version 1.26.0) and Pandas (version 2.1.3) for numerical operations and data handling, while Scikit-learn (version 1.3.1) is utilized for preprocessing and dataset partitioning. Statistical validation and result analysis are carried out using the R programming language (version 4.5.0).

All neural network models are trained using the Adam optimizer with a fixed learning rate of $10^{-3}$ over 20 epochs. The Rectified Linear Unit (ReLU) activation function is employed in all hidden layers to effectively capture non-linear relationships in the structured input space. These training settings are kept consistent across all architectures, while variations are introduced only in the number of hidden layers, the number of neurons per layer, and the batch size, allowing for a controlled investigation of architectural design choices.
All experiments are conducted under fixed random seeds and identical training conditions.

\subsection{Experimental Design}

\subsubsection{Training and Testing Datasets}

The dataset used in this study is generated using BCD. The input variables are sampled from predefined random distributions to ensure sufficient coverage of the system parameter space. For each generated sample, BCD is executed to obtain the corresponding optimal design variables and the achievable data rate, which serve as reference solutions. The training dataset consists of 10,000 samples generated from broad parameter ranges to capture diverse system configurations. Empirically, this dataset size was found sufficient to ensure stable convergence of all considered models, with no significant performance gains observed when increasing the number of training samples further.

To rigorously evaluate the generalizability of CALO, two distinct and unseen testing datasets are constructed, corresponding to \textit{urban} and \textit{rural} deployment scenarios. Each testing dataset contains 2,000 samples and is generated using scenario-specific parameter distributions that reflect realistic propagation environments.

The adopted parameter distributions are designed to reflect both general training conditions and scenario-specific characteristics, capturing variations in propagation distance, power budgets, and channel conditions commonly encountered in RIS-assisted wireless systems, as detailed in Table~\ref{tab:dataset}. The corresponding output variables $(x_1, x_2, x_3, P_1, P_2, N_1, N_2)$ are obtained by solving each instance using BCD, ensuring that all reference solutions satisfy the problem constraints and represent high-quality feasible operating points.

\setlength{\tabcolsep}{1.5pt}
\begin{table}[!t]
\centering
\caption{Parameter distributions for training and testing datasets under different scenarios}
\label{tab:dataset}
\renewcommand{\arraystretch}{1.2}
\scriptsize{
\begin{tabular}{|>{\centering\arraybackslash}p{0.06\columnwidth}|>{\centering\arraybackslash}m{0.15\columnwidth}|>{\centering\arraybackslash}m{0.14\columnwidth}|>{\centering\arraybackslash}m{0.17\columnwidth}|>{\centering\arraybackslash}m{0.16\columnwidth}|>{\centering\arraybackslash}m{0.19\columnwidth}|}
\hline
\textbf{Var.} & \textbf{Description} & \textbf{Type} & \textbf{Training Dataset} & \textbf{Urban Scenario} & \textbf{Rural Scenario} \\
\hline
$D$ & Total horizontal distance (m) & Continuous & $\mathcal{U}(50,200)$ & $\mathcal{U}(20,80)$ & $\mathcal{U}(150,300)$ \\
\hline
$M$ & Number of BS antennas & Discrete & $\mathcal{U}_d(2,17)$ & $\mathcal{U}_d(4,9)$ & $\mathcal{U}_d(2,7)$ \\
\hline
$H$ & Height of RISs (m) & Continuous & $\mathcal{U}(1,6)$ & $\mathcal{U}(2,4)$ & $\mathcal{U}(4,7)$ \\
\hline
$P_F$ & Total RIS power (dBm) & Continuous & $\mathcal{U}(20,28)$ & $\mathcal{U}(20,26)$ & $\mathcal{U}(22,28)$ \\
\hline
$P_B$ & BS transmit power (dBm) & Continuous & $\mathcal{U}(25,35)$ & $\mathcal{U}(28,32)$ & $\mathcal{U}(30,34)$ \\
\hline
$N$ & Total reflecting elements & Discrete & $\mathcal{U}_d(50,1500)$ & $\mathcal{U}_d(50,750)$ & $\mathcal{U}_d(750,2000)$ \\
\hline
$\beta$ & Channel gain (dB) & Continuous & $\mathcal{N}(-30,2)$ & $\mathcal{N}(-32,2)$ & $\mathcal{N}(-28,2)$ \\
\hline
$\delta_1^2$ & Noise power at RIS1 (dBm) & Continuous & $\mathcal{N}(-70,3)$ & $\mathcal{N}(-75,2)$ & $\mathcal{N}(-68,2)$ \\
\hline
$\delta_2^2$ & Noise power at RIS2 (dBm) & Continuous & $\mathcal{N}(-70,3)$ & $\mathcal{N}(-75,2)$ & $\mathcal{N}(-68,2)$ \\
\hline
$\delta_0^2$ & Noise power at user (dBm) & Continuous & $\mathcal{N}(-80,3)$ & $\mathcal{N}(-85,2)$ & $\mathcal{N}(-78,2)$ \\
\hline
\end{tabular}
}
\end{table}

The urban scenario is characterized by shorter transmission distances and more severe propagation conditions, reflected by lower values of $D$ and higher attenuation levels. In contrast, the rural scenario represents more open environments with longer transmission distances and relatively lower attenuation. These distinct distributions allow for evaluating the robustness of the proposed models under heterogeneous deployment conditions. Both testing datasets are generated independently from the training dataset, ensuring that evaluation is performed on unseen data distributions. This design enables a rigorous assessment of the model’s ability to generalize beyond the conditions observed during training and to outperform the BCD baseline under diverse system configurations.

All models are evaluated on identical testing datasets. For each scenario, performance metrics are computed over all samples and include the achievable data rate, the average achievable data rate gain with respect to the BCD baseline, and constraint feasibility. This evaluation protocol enables a systematic assessment of (i) the extent to which the proposed models can surpass the BCD solution, (ii) their ability to preserve feasibility under varying system configurations, and (iii) their generalization capability across heterogeneous deployment environments.

\subsubsection{Evaluation of the Proposed Solution under Different Architectures}

This subsection evaluates the effectiveness of CALO in addressing the underlying non-convex optimization problem. 
Specifically, the evaluation focuses on: (i) the ability to satisfy the problem constraints through learning-based parameterization, (ii) the performance improvement over the BCD solution, and (iii) the computational efficiency of the learned models in enabling fast inference. 
To identify a suitable model architecture, multiple neural network configurations are evaluated by varying the number of layers, neurons per layer, and batch sizes, as summarized in Table~\ref{tab:arch_training}. The considered architectures range from shallow networks with limited capacity to deeper models with a significantly higher number of trainable parameters. For each configuration, both training and validation losses are monitored to assess convergence behavior and generalization capability.
\begin{table}[!t]
\centering
\caption{Training Performance Across Different Architectures (Induction Phase)}
\label{tab:arch_training}
\renewcommand{\arraystretch}{1.2}
\scriptsize{
\begin{tabular}{|>{\centering\arraybackslash}p{0.1\columnwidth}|>{\centering\arraybackslash}m{0.08\columnwidth}|>{\centering\arraybackslash}m{0.08\columnwidth}|>{\centering\arraybackslash}m{0.25\columnwidth}|>{\centering\arraybackslash}m{0.14\columnwidth}|>{\centering\arraybackslash}m{0.14\columnwidth}|}
\hline
\textbf{Layers} & \textbf{Units} & \textbf{Batch} & \textbf{Trainable Params} & \textbf{Train Loss} & \textbf{Val Loss} \\
\hline

\multirow{3}{*}{4} 
& \multirow{3}{*}{16} 
& 64  & \multirow{3}{*}{1111}    & 0.230 & 0.005 \\
& & 128 &                         & 0.448 & 0.267 \\
& & 512 &                         & 1.742 & 1.546 \\
\hline

\multirow{3}{*}{6} 
& \multirow{3}{*}{64} 
& 64  & \multirow{3}{*}{21959}   & 0.075 & 0.003 \\
& & 128 &                         & 0.147 & 0.003 \\
& & 512 &                         & 0.563 & 0.355 \\
\hline

\multirow{3}{*}{8} 
& \multirow{3}{*}{256} 
& 64  & \multirow{3}{*}{465159}  & 0.021 & 0.003 \\
& & 128 &                         & 0.039 & 0.003 \\
& & 512 &                         & 0.144 & 0.006 \\
\hline

\multirow{3}{*}{16} 
& \multirow{3}{*}{512} 
& 64  & \multirow{3}{*}{3949063} & 0.026 & 0.003 \\
& & 128 &                         & 0.048 & 0.003 \\
& & 512 &                         & 0.183 & 0.052 \\
\hline
\end{tabular}
}
\end{table}

Although several architectures achieve comparably low validation loss values, it is observed that validation loss alone is not a sufficient criterion for selecting the best-performing model in this problem. This is primarily due to the nature of the learning objective, where the loss function does not directly capture the true optimization goal, namely the achievable data rate under strict feasibility constraints. As a result, models with similar validation loss may exhibit significantly different performance in terms of constraint satisfaction and rate optimization during testing. Therefore, instead of relying solely on validation loss, the final model selection is based on a comprehensive evaluation that includes feasibility, achievable data rate performance, and computational efficiency. This ensures that the selected architecture not only generalizes well but also effectively solves the underlying non-convex optimization problem. The results for all considered architectures are summarized in Tables~\ref{tab:urban_results} and~\ref{tab:rural_results} for urban and rural scenarios, respectively.
\begin{table}[!t]
\centering
\caption{Performance Evaluation under Different Architectures (Urban Scenario)}
\label{tab:urban_results}
\renewcommand{\arraystretch}{1.2}
\scriptsize{
\begin{tabular}{|>{\centering\arraybackslash}p{0.1\columnwidth}|>{\centering\arraybackslash}m{0.08\columnwidth}|>{\centering\arraybackslash}m{0.08\columnwidth}|>{\centering\arraybackslash}m{0.29\columnwidth}|>{\centering\arraybackslash}m{0.13\columnwidth}|>{\centering\arraybackslash}m{0.19\columnwidth}|}
\hline
\textbf{Layers} & \textbf{Units} & \textbf{Batch} & \textbf{Avg. Inference Time (s)} & \textbf{Feasibility} & \textbf{Avg. Rate Gain} \\
\hline

\multirow{3}{*}{4} & \multirow{3}{*}{16} 
& 64  & 0.00009 & 100\% & 0.425 \\
& & 128 & 0.00009 & 100\% & 0.404 \\
& & 512 & 0.00008 & 100\% & 0.369 \\
\hline

\multirow{3}{*}{6} & \multirow{3}{*}{64} 
& 64  & 0.00010 & 100\% & 0.407 \\
& & 128 & 0.00010 & 100\% & 0.426 \\
& & 512 & 0.00010 & 100\% & 0.383 \\
\hline

\multirow{3}{*}{8} & \multirow{3}{*}{256} 
& 64  & 0.00017 & 100\% & 0.392 \\
& & 128 & 0.00014 & 100\% & 0.426 \\
& & \textbf{512} & \textbf{0.00014} & \textbf{100\%} & \textbf{0.431} \\
\hline

\multirow{3}{*}{16} & \multirow{3}{*}{512} 
& 64  & 0.00030 & 100\% & 0.411 \\
& & 128 & 0.00030 & 100\% & 0.421 \\
& & 512 & 0.00030 & 100\% & 0.423 \\
\hline
\end{tabular}
}
\end{table}

\begin{table}[!t]
\centering
\caption{Performance Evaluation under Different Architectures (Rural Scenario)}
\label{tab:rural_results}
\renewcommand{\arraystretch}{1.2}
\scriptsize{
\begin{tabular}{|>{\centering\arraybackslash}p{0.1\columnwidth}|>{\centering\arraybackslash}m{0.08\columnwidth}|>{\centering\arraybackslash}m{0.08\columnwidth}|>{\centering\arraybackslash}m{0.29\columnwidth}|>{\centering\arraybackslash}m{0.13\columnwidth}|>{\centering\arraybackslash}m{0.19\columnwidth}|}
\hline
\textbf{Layers} & \textbf{Units} & \textbf{Batch} & \textbf{Avg. Inference Time (s)} & \textbf{Feasibility} & \textbf{Avg. Rate Gain} \\
\hline

\multirow{3}{*}{4} & \multirow{3}{*}{16} 
& 64  & 0.00008 & 100\% & 0.312 \\
& & 128 & 0.00008 & 100\% & 0.288 \\
& & 512 & 0.00008 & 100\% & 0.280 \\
\hline

\multirow{3}{*}{6} & \multirow{3}{*}{64} 
& 64  & 0.00009 & 100\% & 0.284 \\
& & 128 & 0.00009 & 100\% & 0.301 \\
& & 512 & 0.00009 & 100\% & 0.292 \\
\hline

\multirow{3}{*}{8} & \multirow{3}{*}{256} 
& 64  & 0.00014 & 100\% & 0.282 \\
& & 128 & 0.00014 & 100\% & 0.309 \\
& & \textbf{512} & \textbf{0.00014} & \textbf{100\%} & \textbf{0.320} \\
\hline

\multirow{3}{*}{16} & \multirow{3}{*}{512} 
& 64  & 0.00029 & 100\% & 0.305 \\
& & 128 & 0.00029 & 100\% & 0.311 \\
& & 512 & 0.00029 & 100\% & 0.295 \\
\hline
\end{tabular}
}
\end{table}

All evaluated models strictly satisfy the problem constraints with zero numerical violation across all test samples. In particular, the equality constraints $\sum_i x_i = D$, $\sum_i N_i = N$, and $\sum_i P_i= P_F$ are inherently enforced through the proposed softmax-based parameterization. 
As shown in Tables~\ref{tab:urban_results} and~\ref{tab:rural_results}, all evaluated architectures consistently achieve a feasibility rate of $100\%$ across both scenarios. 
This property is critical for practical deployment, as infeasible solutions would render the system design invalid regardless of their performance in terms of data rate.

In terms of achievable data rate performance, the results indicate that the architecture with $8$ layers and $256$ neurons, trained with a batch size of $512$, achieves the highest average rate gain in both urban and rural scenarios. Specifically, this configuration yields the best performance despite not exhibiting the lowest validation loss during the training phase. This observation further supports the earlier argument that validation loss alone is not a reliable indicator of the true optimization performance in this problem.

Moreover, the results demonstrate that increasing model complexity beyond a certain point does not necessarily lead to performance improvement. For instance, deeper architectures with significantly higher numbers of parameters do not consistently outperform the selected configuration. This highlights that the proposed learning framework is capable of effectively solving the underlying non-convex optimization problem without requiring excessively deep or computationally expensive models.

From a computational perspective, the proposed learning-based approach offers a significant advantage over the BCD algorithm. The average inference time per sample for the neural network models is on the order of $10^{-4}$ seconds, as shown in Tables~\ref{tab:urban_results} and~\ref{tab:rural_results}. In contrast, the average time required by the BCD algorithm to compute a suboptimal solution is approximately $0.0768$ seconds for the urban scenario and $0.3373$ seconds for the rural scenario.

\subsubsection{Statistical Testing and Comparative Performance}

This subsection compares CALO with the BCD baseline across urban and rural environments using paired achievable-rate differences. Since both methods are evaluated on the same testing samples, the comparison is naturally paired. Therefore, the statistical analysis is performed on the per-sample rate difference
$d^{(i)}=R_{\mathrm{CALO}}^{(i)}- R_{\mathrm{BCD}}^{(i)}$,
where $R_{\mathrm{CALO}}^{(i)}$ and $R_{\mathrm{BCD}}^{(i)}$ denote the achievable rates obtained by CALO and BCD for the \(i\)-th test sample, respectively. This paired formulation directly measures the improvement achieved by CALO over BCD while reducing the effect of sample-to-sample variability.

\begin{figure}[!t]
\centering
\subfloat[Urban scenario.]{
\includegraphics[width=0.4\textwidth]{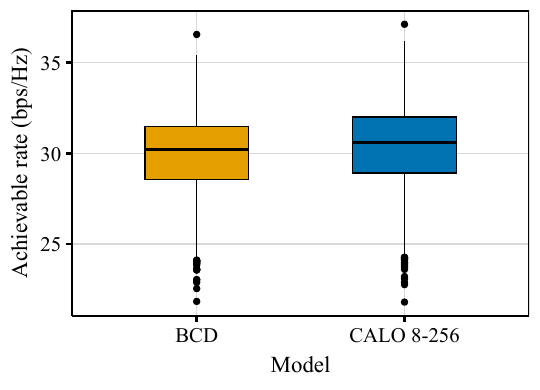}
\label{fig:urban_boxplot}}
\hfill
\subfloat[Rural scenario.]{
\includegraphics[width=0.4\textwidth]{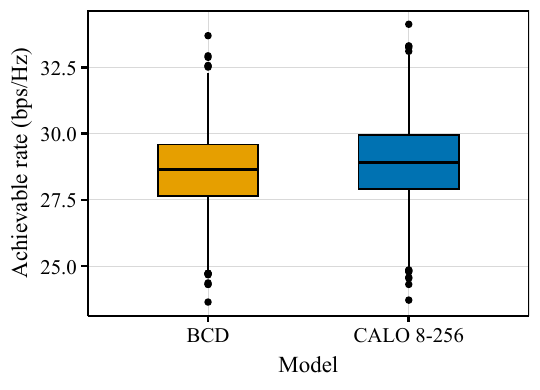}
\label{fig:rural_boxplot}}
\caption{Distribution of the achievable data rate obtained by the BCD baseline and the selected CALO model across the testing samples under (a) urban and (b) rural deployment scenarios.}
\label{fig:boxplot_rate_comparison}
\end{figure}

Figs.~\ref{fig:urban_boxplot} and~\ref{fig:rural_boxplot} show the achievable-rate distributions obtained by CALO and BCD in the urban and rural scenarios, respectively. The boxplots indicate a consistent upward shift of CALO relative to BCD.

The statistical testing procedure is designed to answer three questions: whether a parametric paired test is appropriate, whether the observed improvement is statistically significant, and whether the magnitude of improvement is practically meaningful. First, we apply the Anderson--Darling test to assess the normality of the paired differences, which is a standard goodness-of-fit approach for evaluating distributional assumptions~\cite{dagostino1986goodness}. This step is required because a paired \(t\)-test assumes approximate normality of the paired differences. The Anderson--Darling test rejects normality in both scenarios, with  $p < 2.2\times10^{-16}$. Therefore, we avoid a parametric paired \(t\)-test and use the Wilcoxon signed--rank test, which is a rank-based nonparametric test suitable for paired comparisons when the normality assumption is not satisfied~\cite{conover1999practical,hollander2014nonparametric}.

The Wilcoxon signed--rank test is applied to examine whether CALO provides a positive location shift over BCD. The hypotheses are:
\[
H_0:\ \mathrm{median}(d)=0
\qquad\text{vs.}\qquad
H_1:\ \mathrm{median}(d)>0,
\]
where $d = R_{\mathrm{CALO}} - R_{\mathrm{BCD}}$. For both urban and rural environments, the Wilcoxon signed--rank test yields $p < 2.2\times10^{-16}$, leading to rejection of $H_0$. This result provides strong statistical evidence that CALO achieves a positive paired achievable-rate improvement over BCD.

To complement the \(p\)-values, we report the Hodges--Lehmann (HL) median shift with $95\%$ confidence intervals, which provides a robust estimate of the typical paired improvement~\cite{hollander2014nonparametric}. We also report the standardized Wilcoxon effect size $r = Z/\sqrt{n}$, where $Z$ is the normal approximation of the signed-rank statistic and $n$ is the number of paired samples. Effect-size measures are important because they quantify the strength of the improvement rather than only its statistical significance~\cite{grissom2012effect}. In the urban scenario, the HL shift is $0.4088$ bps/Hz with a $95\%$ confidence interval of $[0.3996, 0.4181]$ bps/Hz. In the rural scenario, the HL shift is $0.3025$ bps/Hz with a $95\%$ confidence interval of $[0.2972, 0.3078]$ bps/Hz. In both cases, the standardized Wilcoxon effect size is very large, with $r \approx 0.8661$. The rank-biserial correlation is approximately one, and the common-language effect size is about $0.9995$, indicating that nearly all paired comparisons favor CALO.

\begin{figure}[!t]
\centering
\subfloat[Urban scenario.]{
\includegraphics[width=0.4\textwidth]{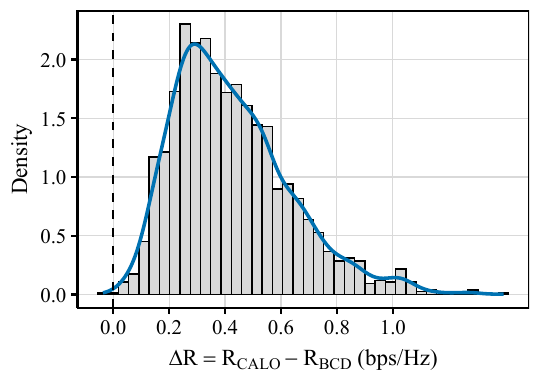}
\label{fig:urban_CALO_histogram}}
\hfill
\subfloat[Rural scenario.]{
\includegraphics[width=0.4\textwidth]{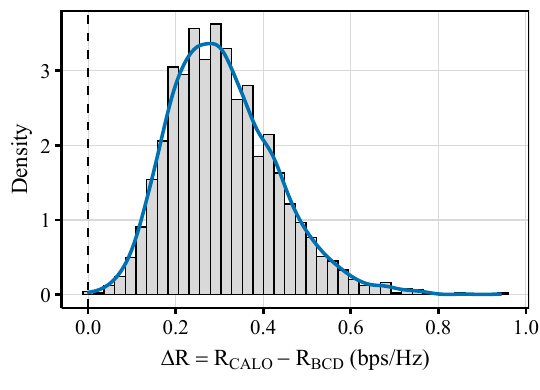}
\label{fig:rural_CALO_histogram}}
\caption{Distribution of the rate difference ($R_{\mathrm{CALO}} - R_{\mathrm{BCD}}$) under urban and rural environments.}
\label{fig:hist_rate_difference}
\end{figure}

Figs.~\ref{fig:urban_CALO_histogram} and~\ref{fig:rural_CALO_histogram} show the distributions of the paired rate differences ($R_{\mathrm{CALO}} - R_{\mathrm{BCD}}$). The distributions are concentrated on positive values, which is consistent with the Wilcoxon test results and confirms that the improvement is not driven by a small number of isolated samples.

Although the relative gains are modest, they are practically meaningful because throughput scales with bandwidth. If $R$ denotes the achievable rate per unit bandwidth, the end-to-end throughput is $T = R \times B$, where $B$ is the system bandwidth. Therefore, an incremental gain $\Delta R$ leads to an absolute throughput increase of $\Delta T = \Delta R \times B$, which can become significant over wide bandwidths and when aggregated across users, time, and slices. To provide a normalized interpretation of the gain, we compute the Mean Relative Improvement (MRI) as:
\[
\text{MRI} = \frac{1}{N} \sum_{i=1}^{N} \left( \frac{R_{\text{CALO}}^{(i)} - R_{\text{BCD}}^{(i)}}{R_{\text{BCD}}^{(i)}} \right) \times 100
\]
The resulting MRI values are approximately $1.29\%$ in the urban scenario and $0.98\%$ in the rural scenario.

Table~\ref{tab:model_comparison} summarizes the statistical and performance results. Overall, the statistical analysis confirms that the observed gains are not due to random fluctuations or isolated test samples. Instead, CALO provides a consistent positive shift over the BCD baseline in both urban and rural environments while maintaining $100\%$ feasibility. Combined with the elimination of per-instance iterative optimization at inference, these results support the main claim that CALO achieves feasible, statistically significant, and practically meaningful performance improvements with real-time applicability.

\begin{table}[!t]
\centering
\caption{CALO vs. BCD across rural and urban environments.}
\label{tab:model_comparison}
\renewcommand{\arraystretch}{1.15}
\resizebox{\columnwidth}{!}{%
\begin{tabular}{lcc}
\hline
\textbf{Metric} & \textbf{Rural} & \textbf{Urban} \\
\hline
Feasibility (\%) & 100 & 100 \\
Avg. rate gain (bps/Hz) & 0.320 & 0.431 \\
Anderson--Darling $p$-value & $<2.2\times10^{-16}$ & $<2.2\times10^{-16}$ \\
Wilcoxon signed--rank $p$-value & $<2.2\times10^{-16}$ & $<2.2\times10^{-16}$ \\
HL shift (bps/Hz) & 0.3025 & 0.4088 \\
Wilcoxon effect size $r$ & 0.8661 & 0.8661 \\
Mean Relative Improvement (MRI) & 0.98\% & 1.29\% \\
\hline
\end{tabular}%
}
\end{table}

\subsection{Performance Analysis Under Varying System Parameters}

To further assess the robustness and generalization capability of CALO, this subsection examines its behavior under varying system parameters. In particular, we evaluate the impact of key design variables, including $D$, $P_F$, and $N$, on the achievable data rate. For each parameter, controlled experiments are conducted by varying the parameter of interest while keeping the remaining variables fixed on the values listed in Table~\ref{tab:system_parameters}, thereby isolating its effect on the system behavior. This analysis provides insights into how the proposed learning-based solution adapts to different operating conditions of the underlying non-convex optimization problem. 

\begin{table}[!t]
\centering
\caption{Default system parameters.}
\label{tab:system_parameters}
\renewcommand{\arraystretch}{0.9}
\resizebox{\columnwidth}{!}{%
\begin{tabular}{cc|cc|cc|cc}
\hline
\textbf{Parameter} & \textbf{Value} & \textbf{Parameter} & \textbf{Value} & \textbf{Parameter} & \textbf{Value} & \textbf{Parameter} & \textbf{Value} \\
\hline
$M$ & $4$ & $H$ & $2$~m & $N$ & $750$ & $D$ & $100$~m\\
\hline
$P_F$ & $26$~dBm & $P_B$ & $30$~dBm & $\beta$ & $-30$~dB & $\delta_0^2$ & $-80$~dBm \\
\hline
$\delta_1^2$ & $-70$~dBm & $\delta_2^2$ & $-70$~dBm \\ \cline{1-4}
\end{tabular}%
}
\end{table}

\subsubsection{Impact of $D$}

To investigate the effect of propagation distance on system performance, $D$ is varied while keeping the remaining system parameters fixed on the values in Table~\ref{tab:system_parameters}.

\begin{figure}[!t]
\centering
\subfloat[Achievable rate vs. distance.]{
    \includegraphics[width=0.4\textwidth]{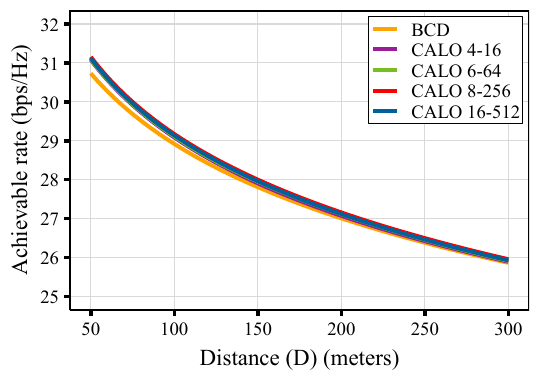}
    \label{fig:rate_vs_distance}}
\hfill
\subfloat[Achievable-rate gain vs. distance.]{
    \includegraphics[width=0.4\textwidth]{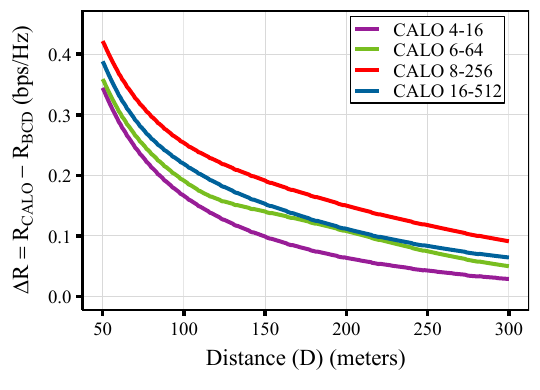}
    \label{fig:gain_vs_distance}}
\caption{Performance versus distance. 
}
\label{fig:distance}
\end{figure}

Fig.~\ref{fig:distance} illustrates the achievable data rate and the corresponding rate gain over the BCD baseline as a function of $D$. As shown in Fig.~\ref{fig:rate_vs_distance}, the achievable data rate decreases monotonically with increasing distance for all considered methods. This behavior is expected due to the increased path loss associated with larger transmitter–receiver separation, which reduces the effective received signal power and consequently the SNR.

Despite this degradation, CALO consistently achieves higher data rates compared to the BCD baseline across the entire range of distances. More importantly, Fig.~\ref{fig:gain_vs_distance} shows that the achievable-rate gain remains positive and relatively stable, indicating that the learned solution maintains its superiority even under unfavorable propagation conditions.

Furthermore, among the evaluated architectures, the configuration with $8$ layers and $256$ neurons demonstrates the most consistent and highest performance gain across all distances. This observation reinforces that an appropriately designed architecture can effectively capture the underlying structure of the non-convex optimization problem without requiring excessive model complexity. Overall, these results confirm that CALO, not only improves performance under nominal conditions but also generalizes well across varying propagation regimes, making it suitable for practical wireless deployment scenarios.

\subsubsection{Impact of $N$}

To evaluate the influence of the number of reflecting elements on system performance, $N$ is varied while keeping the remaining system parameters fixed on the values in Table~\ref{tab:system_parameters}.
\begin{figure}[!t]
\centering
\subfloat[Achievable rate vs. number of reflecting elements.]{
    \includegraphics[width=0.4\textwidth]{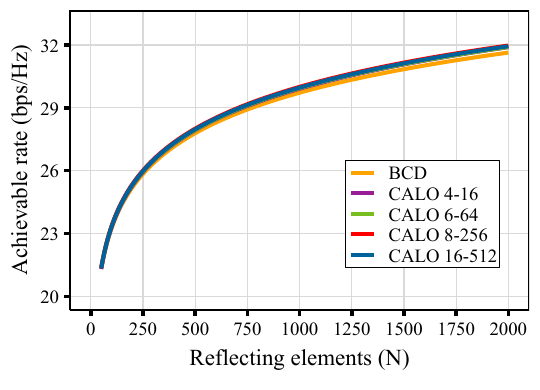}
    \label{fig:rate_vs_n}}
\hfill
\subfloat[Achievable-rate gain vs. number of reflecting elements.]{
    \includegraphics[width=0.4\textwidth]{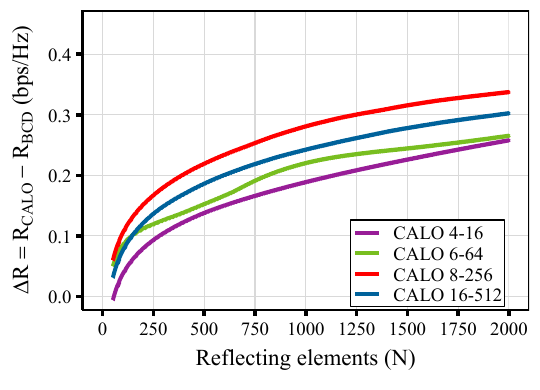}
    \label{fig:gain_vs_n}}
\caption{Performance versus the number of reflecting elements. 
}
\label{fig:reflecting_elements}
\end{figure}

Fig.~\ref{fig:reflecting_elements} presents the achievable data rate and the corresponding rate gain over the BCD baseline as a function of the number of reflecting elements. As shown in Fig.~\ref{fig:rate_vs_n}, the achievable data rate increases with $N$ for all considered methods. This behavior is attributed to the enhanced passive beamforming gain provided by the RIS, where a larger number of reflecting elements enables more precise signal reflection and constructive combining at the receiver, thereby improving the effective channel gain. However, the rate improvement exhibits a diminishing return as $N$ becomes large. This saturation effect is expected, as the incremental contribution of additional reflecting elements decreases once the dominant propagation paths are sufficiently reinforced and the system approaches its performance limits.

Importantly, CALO consistently outperforms the BCD baseline across the entire range of $N$. As illustrated in Fig.~\ref{fig:gain_vs_n}, the achievable-rate gain remains positive and stable, indicating that the learning-based solution effectively captures the relationship between the number of reflecting elements and the optimal resource allocation.

Furthermore, the architecture with $8$ layers and $256$ neurons again demonstrates the most favorable trade-off between performance and model complexity, achieving the highest and most consistent gains across all values of $N$. This result further confirms that the proposed framework does not require excessively deep architectures to approximate high-quality solutions for the underlying non-convex optimization problem.

This behavior indicates that the proposed model effectively learns the nonlinear relationship between RIS size and system performance, which is difficult to capture using traditional optimization methods. Overall, these findings highlight the ability of the proposed approach to generalize across different RIS configurations and to maintain robust performance gains even as the system scales in size.

\subsubsection{Impact of $P_F$}

For the sensitivity analysis of \(P_F\), the total RIS power budget is converted from dBm to watts before evaluating the analytical rate expression.
To investigate the effect of the total RIS power budget, $P_F$ is varied while keeping the remaining system parameters fixed on the values in Table~\ref{tab:system_parameters}.
\begin{figure}
\centering
\subfloat[Achievable rate vs. total RIS power.]{
    \includegraphics[width=0.4\textwidth]{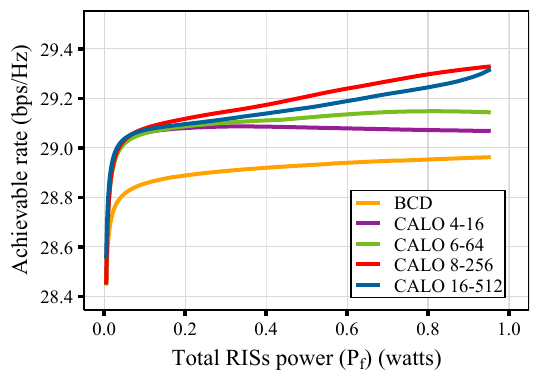}
    \label{fig:rate_vs_pf}}
\hfill
\subfloat[Achievable-rate gain vs. total RIS power.]{
    \includegraphics[width=0.4\textwidth]{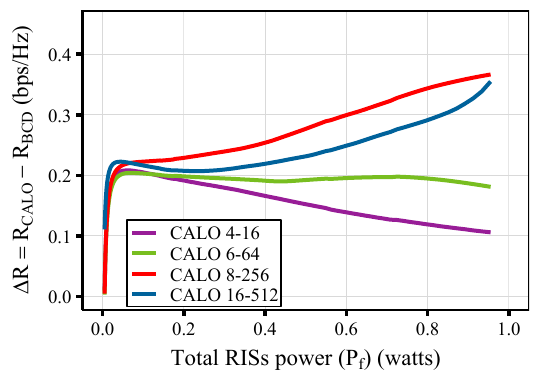}
    \label{fig:gain_vs_pf}}
\caption{Performance versus total RIS power budget. 
}
\label{fig:power_budget}
\end{figure}

Fig.~\ref{fig:power_budget} illustrates the achievable data rate and the corresponding gain over the BCD baseline as a function of the total RIS power. As shown in Fig.~\ref{fig:rate_vs_pf}, the achievable rate increases monotonically with $P_F$ for all considered methods. This behavior is fundamentally different from the passive RIS case, as the active RIS is capable of amplifying the reflected signals, thereby directly enhancing the end-to-end SNR.

Specifically, increasing $P_F$ enables stronger amplification at the RIS elements, which improves the effective cascaded channel gain and compensates for the severe path loss typically associated with multi-hop RIS-assisted communication. As a result, the system transitions from a reflection-limited regime at low $P_F$ to an amplification-assisted regime at higher power budgets. However, similar to the behavior observed with the number of reflecting elements, the rate improvement exhibits a diminishing return as $P_F$ increases. This saturation effect arises because the system becomes progressively constrained by other limiting factors, such as receiver noise and residual interference, reducing the marginal benefit of additional RIS power.

Importantly, CALO consistently outperforms the BCD baseline across the entire range of $P_F$, as shown in Fig.~\ref{fig:gain_vs_pf}. The gain remains positive and relatively stable, indicating that the learning-based model effectively captures the nonlinear coupling between RIS power allocation and system performance. This observation further confirms that optimizing RIS power is a critical factor in active RIS systems, as it directly controls the balance between signal amplification and noise enhancement.

Among all configurations, the architecture with $8$ layers and $256$ neurons again achieves the best performance, demonstrating its ability to accurately model the complex interaction between amplification, resource allocation, and channel conditions without requiring excessive model depth.

These results highlight a key advantage of the proposed approach: its capability to efficiently exploit the additional degrees of freedom introduced by active RISs. In contrast to conventional optimization methods, which may struggle with the increased complexity of power-dependent variables, the proposed learning-based framework maintains robust performance and scalability across different power regimes.

\section{Conclusion}\label{sec:Conclusion}

In this paper, we proposed a constraint-aware learning optimization (CALO) framework for linearly constrained non-convex resource allocation, with a focus on double-active RIS-assisted wireless systems. CALO reformulates the original problem using fractional parameterization over simplex domains, enabling the distance, power, and element-budget constraints to be satisfied by construction. Discrete reflecting-element allocation is handled through a straight-through estimator, while a regret-driven objective allows the model to improve beyond BCD-based reference solutions without relying on globally optimal labels.

Simulation results show that CALO achieves $100\%$ feasibility across all tested configurations without projection or penalty-based corrections. The proposed framework also provides statistically significant achievable-rate gains over the BCD baseline in both urban and rural scenarios, as validated using nonparametric statistical testing. Moreover, by replacing iterative optimization with a single forward pass, CALO reduces online inference time by orders of magnitude, supporting real-time applicability.

Sensitivity analysis under varying propagation distance, number of reflecting elements, and RIS power budget further confirms the robustness and generalization capability of the proposed framework. Overall, the results demonstrate that embedding optimization structure into the learning process can produce feasible, scalable, and high-performance solvers for complex wireless resource-allocation problems.

Future work will extend CALO to more general channel models, multi-user and multi-cell scenarios, and dynamic environments with time-varying system parameters. Incorporating uncertainty-aware and robust learning mechanisms is also an important direction for improving practical deployment.



\bibliographystyle{IEEEtran}
\bibliography{refs_abbr}  
\end{document}